\documentstyle[12pt,psfig]{article}

\newcommand{\ppp}[1]{%
        \setbox0=\hbox{#1}%
        \kern-.02em\copy0\kern-\wd0
        \kern+.04em\copy0\kern-\wd0
        \kern-.02em\raise.0217em\box0}

\newcommand{\lsim}{
 \mathrel{\setbox0=\hbox{$<$}\raise0.6ex\copy0\kern-\wd0
 \lower0.65ex\hbox{$\sim$}}}

\newcommand{\gsim}{
 \mathrel{\setbox0=\hbox{$>$}\raise0.6ex\copy0\kern-\wd0
 \lower0.65ex\hbox{$\sim$}}}

\begin{document}
%
\begin{titlepage}
\renewcommand{\thefootnote}{\fnsymbol{footnote}}
\makebox[2cm]{}\\[-1in]
\begin{flushright}
\begin{tabular}{l}
TUM/T39-97-33
\end{tabular}
\end{flushright}
\vskip0.4cm
\begin{center}
  {\Large\bf  Hard Leptoproduction of Charged Vector Mesons\footnote{Work 
    supported in part by BMBF} }\\ 

\vspace{2cm}
L.\ Mankiewicz\footnote{On leave of absence from N. Copernicus
Astronomical Center, Polish Academy of Science, ul. Bartycka 18,
PL--00-716 Warsaw (Poland)}, 
G. Piller and T. Weigl

\vspace{1.5cm}

\begin{center}
{\em Physik Department, Technische Universit\"{a}t M\"{u}nchen, \\
D-85747 Garching, Germany} 
\end{center}

\vspace{3cm}


\vspace{3cm}

\centerline{\bf Abstract}
\begin{center}
\begin{minipage}{15cm}
We present an analysis of twist-$2$, leading order QCD amplitudes for hard
exclusive leptoproduction of charged vector mesons. 
These processes are determined by  nonforward parton distribution 
functions which are  nondiagonal in quark flavor.
We derive relations between flavor diagonal and nondiagonal
distribution functions based on isospin symmetry.  
Furthermore, we discuss general features of   $\rho^+$ and $\rho^-$ 
production cross sections, and present estimates based 
on a simple model for  nonforward distribution functions. 
\end{minipage}
\end{center}

\end{center}
\end{titlepage}
\setcounter{footnote}{0}

\newpage

\section{Introduction}

Recently much effort has been devoted to study properties of 
nonforward parton distribution functions.  
They can be viewed as  generalizations of ordinary parton distributions.  
In addition they  are  closely related to nucleon form factors. 
Thus, they combine different aspects of the nucleon 
structure and offer new insights.
Although being discussed already some time ago 
\cite{Pioneer,Leipzig,Pioneer2}, 
new detailed investigations of their properties 
were initiated recently in the context 
of the  nucleon spin structure \cite{Ji96,Rad97}. 

To summarize the relation of nonforward to ordinary parton
distributions, recall that the latter can be
represented at the twist-$2$ level as normalized Fourier 
transforms of forward nucleon matrix
elements of nonlocal QCD operators which are constructed
as gauge-invariant overlap of two quark or gluon fields separated by a
light-like distance \cite{Collins,BalBra}.  
Nonforward parton distributions are defined by
the same nonlocal operators -- just sandwiched between nucleon
states with different momenta and eventually spin.  Equally close is
the relation to nucleon form factors which are defined by the
nonforward matrix elements taken in the local limit.

Nonforward parton distributions are probed in processes where the
nucleon target recoils elastically.  To select twist-$2$ correlations
a large scale has to be involved.  One possible process is
deeply-virtual Compton scattering \cite{Ji96,Rad97}.  Another
promising class of reactions sensitive to nonforward distribution
functions is exclusive hard meson production as discussed initially in
Ref.\cite{Rad97}.  
In a recent publication \cite{MPW97} we have studied the
production of neutral mesons.  In this work we focus on charged vector
meson production. 
Here new distribution functions are probed which are nondiagonal in flavor. 
They describe a situation where, for example, 
an up-quark from a proton is removed and a
down-quark is returned to form a neutron.  In this letter
we derive relations between flavor diagonal and nondiagonal
distribution functions based on isospin symmetry.  Subsequently, we
use these relations to estimate $\rho^+$ and $\rho^-$
production cross sections within a simple model for the involved
nonforward distribution functions.

Data on the exclusive production of neutral vector mesons have  
been taken lately at high center of mass energies at HERA 
(for a review and references see Ref.\cite{Crit97}). 
In the measured kinematic domain the corresponding production 
cross sections are controlled by the nonforward gluon 
distribution of the target. 
This will be quite different in charged vector meson production 
where gluon contributions are absent in leading order.

This letter is organized as follows: in Sec.2 we present the 
production amplitude for charged vector meson production. 
Relations between flavor diagonal and nondiagonal 
distribution functions are derived in Sec.3. 
In Sec.4 we discuss results for $\rho^+$ and $\rho^-$ production. 
Finally we give a short summary.

\section{Production amplitude} 
       
Solid QCD descriptions of hard exclusive meson production processes 
are based on the factorization of long- and short-distance dynamics 
which has been proven recently in Ref.\cite{CFS97}. 
In this work it has been shown that 
at large photon virtualities $Q^2 \gg \Lambda_{QCD}^2$ and 
moderate momentum transfers $|t| \sim \Lambda_{QCD}$ 
the amplitudes for the production of 
mesons from longitudinally polarized photons can be split into three parts: 
the perturbatively calculable hard photon-parton interaction arises 
from short
distances, while the long-distance dynamics can be absorbed  
in nonperturbative meson distribution amplitudes and nucleon nonforward 
parton distributions. 
In \cite{MPW97} we have outlined the derivation of the amplitudes 
for neutral meson production. Amplitudes for charged vector meson 
production can be obtained in a similar way.
In leading order in the strong coupling constant $\alpha_s$ 
we obtain for the $\rho^+$-production amplitude: 
\begin{eqnarray} 
&& \hspace*{-1cm} {\cal A}^{\rho^+} = \, \pi \, \alpha_s \, 
           \frac{C_F}{N_c} \frac{1}{Q} 
           \frac{\bar{N}(P',S') \hat{n} N(P,S)}{{\bar P} \cdot n}
           f_{\rho}^L
           \int_0^1 d\tau \frac{\Phi_{\rho}^L(\tau)}{\tau \bar{\tau}}
          \int_0^1 dx \int_0^{\bar{x}} dy \, 
\nonumber \\
   &\times & \left[ (e_d\, F^{ud} + e_u\, {\bar F}^{ud}) 
   \frac{{\bar \omega}}{x + 2 y + x \bar{\omega} - i \epsilon} - 
    (e_u \, F^{ud} + e_d\, {\bar F}^{ud})
    \frac{{\bar \omega}}{x + 2 y - x \bar{\omega} - i \epsilon} \right]  
    \nonumber \\
   &+& K-\mbox{terms}.
\label{eq:vecm_quark}
\end{eqnarray}
$N(P,S)$ ($N(P',S')$) stands for the  Dirac spinor  
of the initial (scattered) nucleon with four momentum $P$ ($P'$) 
and spin $S$ ($S'$). 
Here  and in the following we do not explicitly
specify the contribution arising from the so-called ``$K$-terms'' 
which are proportional to the momentum 
transfer $r = P - P'$ (see e.g. \cite{Ji96,Rad97,MPW97}). 
The average nucleon momentum is denoted by $\bar P = (P + P')/2$. 
The four momentum of the incident virtual photon is 
$q$ and $Q^2 = - q^2$. 
The produced vector meson carries the four momentum 
$q'$ and $\bar q = (q + q')/2$. 
In addition we have introduced the variable 
$\bar \omega = 2 \bar q \cdot \bar P/ (- \bar q^2)$. 
Furthermore, $n$ denotes a light-like vector with 
$n\cdot a = a^+ = a^0+ a^3$ \
for any vector $a$, and $\hat n = \gamma_{\mu} n^{\mu}$. 
In Eq.(\ref{eq:vecm_quark}) the perturbative photon-parton 
interaction has been calculated to  leading twist accuracy.
As compared to the virtuality of the photon $Q^2$ we have therefore 
neglected  
the momentum transfer $t= (P-P')^2$, the invariant mass of the 
nucleon target $P^2 = P'^2$, and the mass of the 
produced vector meson $q^{'2}$.

All information about the long-distance dynamics of the 
produced longitudinally polarized $\rho$ meson is contained in the  
decay constant $f_{\rho}^L$ and the distribution amplitude   
$\Phi_{\rho}^L$  \cite{BallBraun}:  
\begin{equation} \label{eq:meson_distr_VL}
  \langle \rho^\pm (q') |\, \bar{\psi}(x) \hat n \psi(y) \, | 0 \rangle
  =  \pm q'\cdot n \,f_{\rho}^L \int_0^1 d\tau \, 
    \Phi_{\rho}^L(\tau) \,
    e^{i q' \cdot \, (\tau x + \bar{\tau} y)}. 
 \end{equation} 
Here the quark fields $\psi$ are understood to carry 
proper flavor quantum numbers.

Contrary to neutral meson production  \cite{MPW97} 
the nucleon part of the amplitude in Eq.(\ref{eq:vecm_quark}) 
is determined by  nucleon double distribution functions $F^{ud}$ and 
$\bar F^{ud}$  which are nondiagonal in flavor. 
They correspond to a situation where e.g., an up-quark with 
electromagnetic charge $e_u$ 
is removed from the target but a down-quark with charge $e_d$ is returned. 
The operator definition of these genuine new object reads:
\begin{eqnarray} \label{eq:dd_quark_ud}
&&\hspace{-0.5cm}
\left \langle p(P',S')\right| 
\bar\psi_u(0) \hat n \left[0,z\right] \psi_d(z) 
\left|n(P,S)\right \rangle_{z^2 = 0} 
\!=\!
\bar N(P',S')\,\hat n \,N(P,S)  
\nonumber \\
&&\times
\!\int_0^1 \!\!dx \!\int_0^{\bar x} \!\! dy\! 
\left[
e^{-i x (P\cdot z) - i y(r\cdot z)} F^{ud}(x,y,t) 
- 
e^{i x (P\cdot z) - i \bar y(r\cdot z)} \bar F^{ud}(x,y,t)
\right] 
+ K-\mbox{terms}.  
\nonumber \\
\end{eqnarray}
Here $\psi_u$ and $\psi_d$ represent up- and down-quark fields which 
are separated by a light-like distance $z\sim n$. 
The proton and neutron states are denoted by 
$\left|p(P',S')\right \rangle$ and $\left|n(P,S)\right \rangle$.
Gauge invariance is guaranteed by the 
path-ordered exponential  $[0,z] = \-{\cal P} \mbox{exp} [ -i g z_{\mu} 
\-\int_0^1 d\lambda A^{\mu}(z \lambda)]$ which reduces to 
one in axial gauge $n\cdot A=0$ 
($g$ stands for the strong coupling constant and 
$A^{\mu}$ denotes the gluon field).

The amplitude for $\rho^-$-production
can be obtained from Eq.(\ref{eq:vecm_quark})
by exchanging quark flavors, $u \leftrightarrow d$, and applying an
overall minus sign due to the definition of the corresponding distribution 
amplitude (\ref{eq:meson_distr_VL}). 

Note that due to the necessary charge transfer in the t-channel, 
gluons do not contribute in leading order 
to  hard exclusive leptoproduction of 
charged mesons.

\section{Isospin relations}

In the following we use isospin symmetry to relate  flavor
nondiagonal double distribution functions  to flavor diagonal ones. 
Because the argument of the nonlocal twist-2 string operator 
in Eq.(\ref{eq:dd_quark_ud}) is a light-like vector, 
it is convenient to introduce conserved isospin charges defined on a
light-like hyper-surface:
\begin{equation}
{\hat \tau}^i = \frac{1}{2} \int d^2 x^\perp dx^- {\bar \Psi}(x^+=0,{\overline 
x}) \gamma^+ \tau^i \Psi(x^+=0,{\overline x}),\
\label{eq:iso_charge}
\end{equation}
where ${\overline x} = (x^-,x^\perp)$, $\Psi = (\psi_u,\psi_d)^{\rm
T}$, and $\tau^i$ are common Pauli matrices.  Since the definition of
${\hat \tau}^i$ involves only ''good'' components of the spinor
fields, canonical anti-commutation relations on the light-cone can be
used \cite{KoSop}.  They lead to:
\begin{equation}
[{\hat \tau}^i,{\hat \tau}^j] = i \epsilon^{ijk}{\hat \tau}^k. 
\end{equation}
Hence the charges in Eq.(\ref{eq:iso_charge}) are indeed generators of 
the isospin symmetry. 
As a consequence in an isospin symmetric world proton and neutron states can 
be related to each other by the usual ladder operators 
${\hat \tau}^\pm = {\hat \tau}^x \pm i {\hat \tau}^y$: 
\begin{eqnarray}
&&{\hat \tau}^+ \left|n\right \rangle = \left|p \right \rangle, 
\quad
{\hat \tau}^- \left|p\right \rangle = \left|n\right \rangle,
\nonumber \\
&&{\hat \tau}^+ \left|p\right \rangle  =  0, 
\quad \hspace*{0.4cm}
{\hat \tau}^- \left|n\right \rangle  = 0.
\label{eq:tau_action}
\end{eqnarray}

As a next step we consider the nonforward matrix element  
(\ref{eq:dd_quark_ud}) which defines the flavor nondiagonal 
double distribution function $F^{ud}$ and $\bar F^{ud}$.
Using the notation 
\begin{equation}
\left.{\hat O}^{q\,q^\prime}(z) =  \bar\psi_q(0) \hat n \left[0,z\right]
\psi_{q^\prime}(z)\right|_{z^2 = 0}, 
\end{equation}
with quark flavors $q$ and $q^\prime$ 
one obtains from Eq.(\ref{eq:tau_action}):
\begin{eqnarray}
\left \langle p\right| {\hat O^{ud}}(z) \left|n\right \rangle &=&   
\left \langle p\right| 
{\hat O^{ud}}(z) \, {\tau}^-\left|p\right \rangle 
= 
\left \langle p \right| 
[{\hat O^{ud}}(z),{\hat \tau}^-]
\left|p \right \rangle,  \nonumber \\ 
&=& 
\left \langle p\right| {\hat O^{uu}}(z) \left|p\right \rangle - 
\left \langle p \right|  {\hat O^{dd}}(z) \left|p\right \rangle,
\label{eq:iso_rel_1}
\end{eqnarray}
and, similarly:
\begin{eqnarray}
&&
\left \langle p\right| {\hat O^{ud}}(z) \left|n\right \rangle =  
\left \langle n\right| {\hat O^{dd}}(z) \left|n\right \rangle - 
\left \langle n\right| {\hat O^{uu}}(z) \left|n\right \rangle, \nonumber \\
&& 
\left \langle n \right| {\hat O^{du}}(z) \left|p\right \rangle =   
\left \langle p\right| {\hat O^{uu}}(z) \left|p\right \rangle - 
\left \langle p\right| {\hat O^{dd}}(z) \left|p\right \rangle, \nonumber \\
&& 
\left \langle n\right| {\hat O^{du}}(z)\left|p\right \rangle =    
\left \langle n\right| {\hat O^{dd}}(z) \left|n\right \rangle - 
\left \langle n\right| {\hat O^{uu}}(z) \left|n\right \rangle. 
\label{eq:iso_rel_2}
\end{eqnarray}

These relations 
enable us to express the charged meson production amplitudes 
in Eq.(\ref{eq:vecm_quark}) in terms of flavor diagonal double distribution 
functions $F^{u} \equiv F^{uu}$ etc.:
\begin{eqnarray}
&& {\cal A}^{\rho^{\pm}} \sim \int_0^1 dx \int_0^{\bar{x}} dy \,
   \left[(F^u - {\bar F}^u) -  (F^d - {\bar F}^d)\right] 
 \left(  \frac{{\bar \omega}}{x + 2 y + x \bar{\omega} - i \epsilon} +
         \frac{{\bar \omega}}{x + 2 y - x \bar{\omega} - i \epsilon} \right)
\nonumber \\
&& \mp \frac{1}{3}
\int_0^1 dx \int_0^{\bar{x}} dy \,
   \left[(F^u + {\bar F}^u) -  (F^d + {\bar F}^d)\right] 
 \left(\frac{{\bar \omega}}{x + 2 y + x \bar{\omega} - i \epsilon} -
         \frac{{\bar \omega}}{x + 2 y - x \bar{\omega} - i \epsilon} \right) 
\nonumber \\
&& +  K-\mbox{terms}.
\label{eq:vecm_quark_diag}
\end{eqnarray}
Here we have inserted the electromagnetic charges 
$e_u+e_d = \frac{1}{3}$, and $e_u-e_d = 1$. 
Note that the second term in Eq.(\ref{eq:vecm_quark_diag}) 
can be related to the isovector part of the $\rho^0$ production 
amplitude from Ref.\cite{MPW97}, 
${\cal A}^{\rho^0}_{(I = 1)} = \frac{1}{2 \sqrt{2}} 
({\cal A}^{\rho^+} - {\cal A}^{\rho^-})$.

\section{Model} 
\label{results}

In the following we discuss the production cross sections for 
$\rho^{+}$ and $\rho^-$ production using  model distribution 
functions introduced recently in Ref.\cite{MPW97}. 
For flavor diagonal distributions we choose \cite{private}:  
\begin{equation} 
  F(x,y,t;\mu_0^2)  =  h(x,y) \, f(x,\mu_0^2)  \,  f(t) ,
\label{def:nf_part_dist}
\end{equation}
where $f(x,\mu_0^2)$ stands for the corresponding ordinary quark 
distribution which we take from Ref.\cite{CTEQ4}.   
As shown in Ref.\cite{MPW97}, double distribution functions 
have to be symmetric with respect to an exchange of variables,   
$y \leftrightarrow 1 - y - x$. 
In accordance with this constraint we use:
\begin{equation}
   h(x,y)  = 6 \,\frac{ y \, (1 - x - y)}{(1-x)^3} 
\label{poss}.
\label{hdef}
\end{equation}
The form factor $f(t)$ 
is responsible for the $t$-dependence of  double distributions.
Motivated by the relationship between double distributions and
nucleon form factors we assume  $f(t) = e^{b t/2}$ 
with $b = 5\,{\rm GeV}^{-2}$ (see e.g. discussion in \cite{MPW97}):

For the $\rho$ meson distribution amplitude we use the asymptotic 
form \cite{Rad,BroLep}: 
\begin{equation}
\Phi_{\rho}(\tau) = 6 \tau (1-\tau).  
\end{equation}
Finally we fix the $\rho$ meson decay constant 
$f_\rho^L = 195$ MeV \cite{BallBraun}.  

In our numerical analysis we neglect the practically 
unconstrained "$K$-terms". As their contribution enters proportional to 
the momentum transfer they are bound to be small at small $r$.

In Fig.1 we present the $\rho^+$ and $\rho^-$ total production 
cross sections from a proton and neutron target, respectively, 
for $Q^2 = 10\,{\rm GeV}^2$.
At small values of $x_{Bj}$ the cross sections 
are approximately proportional to $x_{Bj}$. 
This results from the small-$x_{Bj}$ behavior of the 
flavor nonsinglet combination of ordinary 
quark distributions which enter in 
Eqs.(\ref{eq:vecm_quark_diag},\ref{def:nf_part_dist}). 
At large $x_{Bj}$ the production cross sections drop 
with rising $x_{Bj}$.

It is important to note that the shape of the production 
cross sections  can be motivated from quite general grounds,  
applicable beyond our leading order, leading twist treatment.
The rise at small $x_{Bj}$  
with increasing $x_{Bj}$ is expected from the exchange 
of the leading reggeon poles $\rho$ and $a_2$ \cite{Col77}. 
On the other hand at large values of $x_{Bj}$ the 
minimal momentum transfer to the nucleon target 
which is required for the production of a vector meson, 
$t_{min} \approx - x_{Bj}^2 M^2/(1-x_{Bj})$, 
becomes of the order of typical nucleon scales.
For example at $x_{Bj} = 0.5$ one has $-t_{min} \approx 0.5$ GeV$^2$. 
For such values of $t$ elastic production processes 
from nucleons are suppressed. 
A decrease of $\rho^+$ and $\rho^-$ production 
at large $x_{Bj}$ is the consequence.

Since electromagnetic processes are not isospin invariant 
the production cross sections for $\rho^+$ and $\rho^-$ may differ. 
In Fig.2 we show the ratio of both cross sections.
In the kinematic region $x_{Bj} \sim 0.3$, where both cross sections  
are relatively large, 
$\rho^+$ production from a proton is around $30 \%$ 
larger than $\rho^-$ production from a neutron. 
At small-$x_{Bj}$ both cross sections are of equal size. 
Also this feature is quite general and is due to the  approximate flavor
symmetric sea ${\bar F}^u(x,y) \approx {\bar F}^d(x,y)$, and the small-$x$ 
behavior $F^u(x,y) - {\bar F}^u(x,y) \sim x^{-0.5}$ 
which is expected from Regge theory \cite{Col77}. 

Finally we should  mention that a QCD study of neutral 
vector meson leptoproduction processes emphasizes the role 
of higher twist effects \cite{FS96,Zhy97}. 
A systematic analysis of their importance, also in charged 
vector meson production, is needed.

\section{Summary}

We have discussed hard exclusive leptoproduction of charged vector mesons 
at twist-$2$ leading order $\alpha_s$ accuracy. 
In this framework the considered reactions 
are determined by  nonforward parton distribution functions which 
are nondiagonal in quark flavor. 
Through isospin symmetry these are related to flavor diagonal 
distribution functions.  
Within a simple model for flavor diagonal double distribution functions 
estimates for  $\rho^+$ and $\rho^-$ production  have been given. 
Independent of specific model assumptions we find 
that the corresponding production cross sections are peaked at 
moderate values of $x_{Bj}$. 
Furthermore, at small $x_{Bj}$ both cross sections are of similar size. 
Differences are expected to show up at moderate $x_{Bj}$ 
where the $\rho^+$ and $\rho^-$  cross sections assume their maximum.

\bigskip
\bigskip

\noindent
{\bf Acknowledgments}: 
This work was supported in part by BMBF and KBN grant 
\linebreak 
2~P03B~065~10.

\newpage


\newpage


\begin{figure}[t]
\centering{\psfig{figure=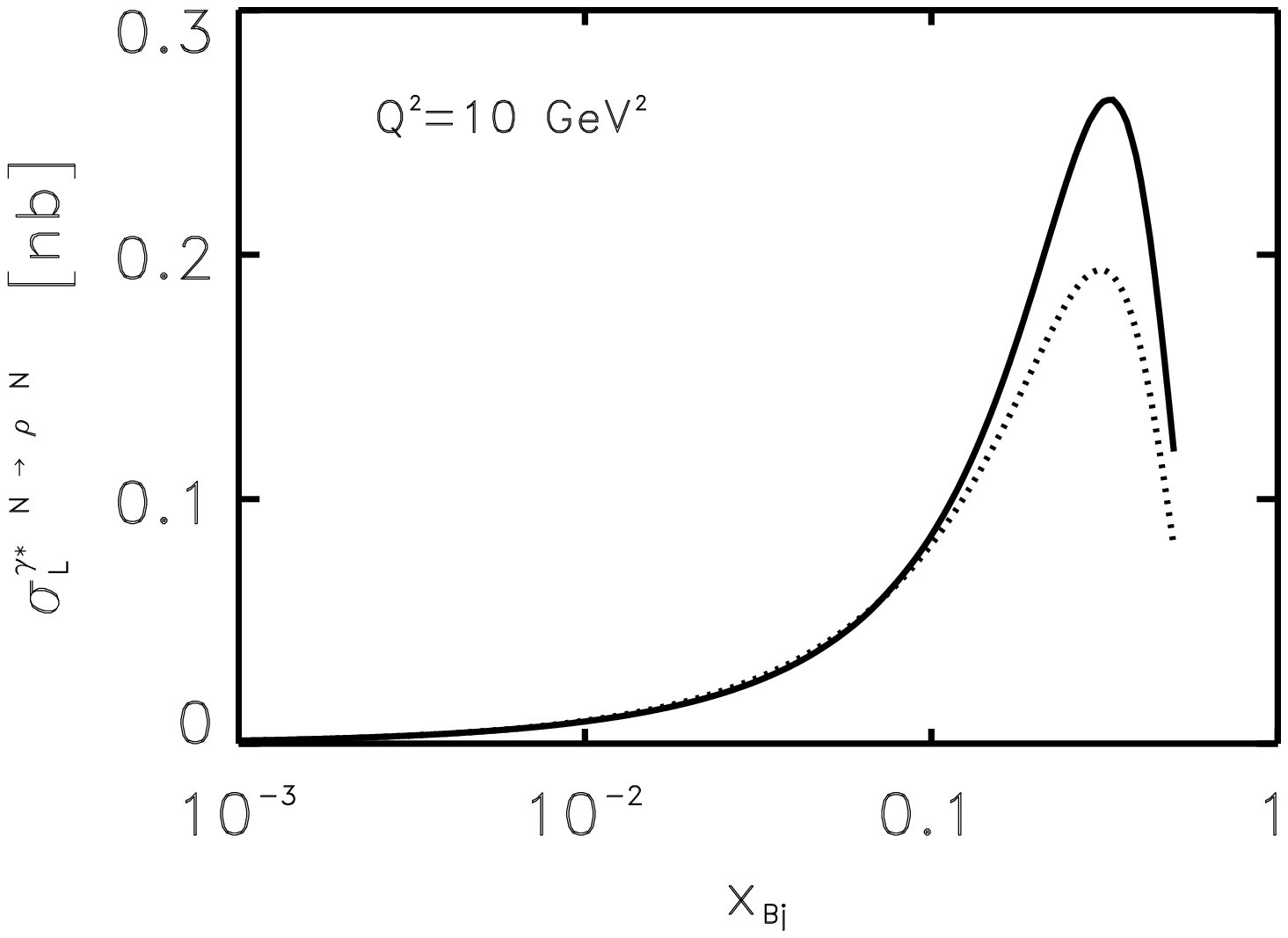,height=8.5cm}}
\caption{Total cross sections for $\rho^+$ (full) and $\rho^-$ (dashed) 
production from a nucleon through the interaction of a longitudinally 
polarized photon.}
\label{fig:cross_section}
\end{figure}

\begin{figure}[b]
\centering{\psfig{figure=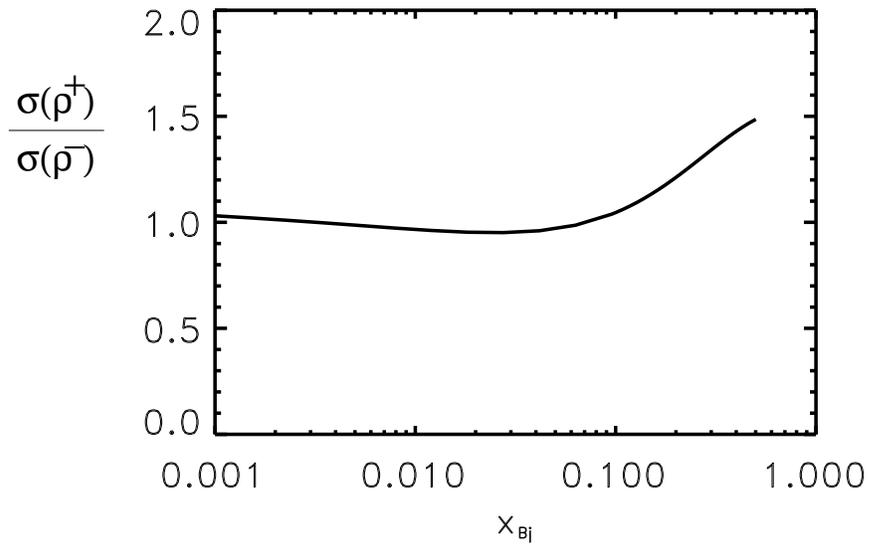,height=8.5cm,angle=-90}}
\caption{Ratio of cross sections for $\rho^+$ and $\rho^-$ production 
from a nucleon through the interaction of a longitudinally polarized photon.}
\label{fig:cross_section_ratio}
\end{figure}

\end{document}